\documentstyle[preprint,aps,eqsecnum]{revtex}
 
\def\beq{\begin{equation}}
\def\eeq{\end{equation}}
\def\rarrow{\rightarrow}

\newcommand{\bea}{\begin{eqnarray}}
\newcommand{\eea}{\end{eqnarray}}
\def\bra#1{\left\langle #1\right\vert}
\def\ket#1{\left\vert #1\right\rangle}

\begin{document}
{
\tighten

\title{Theoretical Analysis of Static Hyperon Data for HYPERON99} 
 
\author{Harry J. Lipkin\,$^{a,b,c}$}
 
\address{ \vbox{\vskip 0.truecm}
  $^a\;$Department of Particle Physics \\
  Weizmann Institute of Science, Rehovot 76100, Israel \\
\vbox{\vskip 0.truecm}
$^b\;$School of Physics and Astronomy \\
Raymond and Beverly Sackler Faculty of Exact Sciences \\
Tel Aviv University, Tel Aviv, Israel \\
\vbox{\vskip 0.truecm}
$^c\;$Department of Physics, U46 \\
  University of Connecticut, 2152 Hillside Rd., Storrs, CT 06269-3046} 

\maketitle

\begin{abstract} 

We consider all hyperon data relevant to spin and flavor structure of 
hyperons. In addition to  masses and magnetic moments considered as static 
properties in Hyperon99 we include also relevant data from hyperon
decays and spin structure determined from deep inelastic scattering.  
Any theoretical model for the hyperons
with parameters to be determined from experiment should use input from all
these data.
Of particular interest are new data from $\Xi^o$ decay and the polarixation
of $\Lambda's$ produced in $Z^o$ decays and deep inelastic scattering.
\end{abstract}
} 

\section{Introduction - What is Meant by Static Properties}

In the context of this meeting, hyperon masses and magnetic moments are 
considered static properties to be discussed in this talk, while hyperon
decays and spin structure determined from deep inelastic scattering are 
not considered static properties. But all of them depend upon the spin and
flavor structure of the hyperons. Any theoretical model for the hyperons
with parameters to be determined from experiment should use input from all
these data.

The masses and magnetic moments are very well measured, and they are well
described by the simple constituent quark model. Going beyond this model
is difficult without  input from other data, some of which are not so well
measured. Thus progress in understanding hyperon structure will come from
combining input from all relevant experimental data and improvements in 
the precision of data other than masses and magnetic moments.
 
\subsection{New Data on $\Xi^o \rightarrow \Sigma^+$ decays}

The new data for the semileptonic decay $\Xi^o \rightarrow \Sigma^+$ 
agrees with the SU(3) prediction 
\beq
g_A(\Xi^o \rightarrow \Sigma^+) =  g_A(n\rightarrow p)     
\eeq
where we use the shortened form $g_A$ to denote  $G_A/G_V$.
 
The essential physics of this prediction is that the spin physics in the 
nucleon system which is probed in the neutron decay is unchanged when the
$d$ quarks in the nucleon are changed to strange quarks. This very
striking result is completely independent of any fitting of weak 
decays using the conventional D/F parametrization. We can immediately
carry this physics further by inserting the $d \leftrightarrow s$
transformation into the well-known prediction for the ratio of the proton
($uud$) to neutron ($udd$) magnetic moments and obtain a prediction for the 
ratio of the $\Sigma^+ (uus)$ to $\Xi^o (uss)$ 
magnetic moments. It is conventient to write the prediction in the form:
\beq
{{\mu_p}\over{\mu_n}} = -1.46 =
{{4 \xi_{ud} + 1}\over{4 + \xi_{ud}}} = -1.5
 \label{QQ1.7a}
\eeq
where $\xi_{ud}$ is the ratio of the quark magnetic moments
\beq
\xi_{ud} = \mu_u/\mu_d = -2
\eeq

This is easily generalized to give 
\beq
{{\mu_\Sigma^+}\over{\mu_\Xi^o}} = -1.96 =
{{4 \xi_{us} + 1}\over{4 + \xi_{us}}} = -1.89
 \label{QQ1.7b}
\eeq
where $\xi_{us}$ is the ratio of the quark magnetic moments, 

\beq
\xi_{us} = \mu_u/\mu_s = (\mu_u/\mu_d)\cdot (\mu_d/\mu_s) = -3.11
\eeq
and we have determined $(\mu_d/\mu_s)$ by the ratio between the experimental 
value of $\mu_\Lambda$ and the SU(3) prediction $\mu_\Lambda =\mu_n/2$
which assumes that $\mu_d = \mu_s$ 
\beq
\mu_d/\mu_s = \mu_n/2\mu_\Lambda = -3.11
\eeq

The fact that the prediction for this ratio (\ref{QQ1.7b}) agrees with
experiment much better than either moment agrees with the SU(6) quark
model\cite {Cooper} is very interesting. 
   
\subsection{New $\Lambda$ polarization measurements from Z decays and DIS} 

When a $\Lambda$ is produced either from $Z^o$ decay or in deep inelastic
scattering, the accepted mechanism is the production of a polarized quark
produced in a pointlike vertex from a $W$ boson or a photon, and the eventual
fragmentation of this quark into the $\Lambda$ directly or into a $\Sigma^o$ or
$\Sigma^*$ which eventually decays into a $\Lambda$. Now that experimental data
on $\Lambda$ polarization are becoming available in both
processes\cite{Lambdapol,e665,LEP-Lambdas,HERA-Lambdas,herm} a central 
theoretical question is which model to use for the spin structure of the
$\Lambda$. In the simple quark model, the strange quark carries the spin of the
$\Lambda$ and the $u$ and $d$ are coupled to spin zero. This model has been
used\cite {gus}  in the first analysis of experimental data from $Z$
decay\cite{LEP-Lambdas} and found to be consistent with the data. But the deep
inelastic experiments have shown that the spin structure of the proton is
different from that given by the simple quark model. An alternative approach is
presented in \cite{jaf} where SU(3) symmetry is assumed and the spin structure
of SU(3) octet hyperons is deduced from that of the proton. But SU(3) symmetry
is known to be broken. Several approaches to this symmetry breaking have been
proposed by theorists\cite{Lambdapol,numagmom}, and other mechanisms are
discussed in \cite{def,thom,ekks,elliskk}. How to include the $\Lambda$'s 
produced by the fragmentation of a quark into $\Sigma^*$ which eventually 
decays into a $\Lambda$ remains controversial, since \cite{Lambdapol,jaf} the
decay via a strong interaction may be already included in the fragmentation 
function. The question of how to all this right remains open. 

\section{Masses and Magnetic Moments}

\subsection{The Sakharov-Zeldovich 1966 Quark model (SZ66)}

Andrei Sakharov was a pioneer in hadron physics who
took quarks seriously already in 1966.
He asked ``Why are the $\Lambda$ and $\Sigma$ masses
different? They are made of the same quarks!"\cite{Sakhaut}.
His answer that the difference arose from
a flavor-dependent hyperfine interaction led to relations
between meson and baryon masses in surprising agreement with
experiment\cite{SakhZel}.
Sakharov and Zeldovich $anticipated$ QCD
by assuming a 
  quark model for hadrons with a flavor dependent linear mass
term and hyperfine interaction,
\begin{equation}
M = \sum_i m_i + \sum_{i>j} 
{{\vec{\sigma}_i\cdot\vec{\sigma}_j}\over{m_i\cdot m_j}}\cdot v^{hyp}_{ij} 
\end{equation}
where $m_i$ is the effective mass of quark $i$, $\vec{\sigma}_i$ is a quark 
spin operator and $v^{hyp}_{ij}$ is
a hyperfine interaction with different strengths
but the same flavor dependence
for $qq$ and $\bar q q$ interactions.

Hadron magnetic moments are are described simply by
adding the contributions of the moments of these constituent quarks with Dirac
magnetic moments having a scale determined by the same  effective masses. The
model describes low-lying excitations of a complex system with remarkable
success. 

Sakarov and Zeldovich already in 1966 obtained two relations between meson and
baryon masses in remarkable agreement with experiment.
Both the mass difference $m_s-m_u$ between strange and nonstrange quarks and
their mass ratio $m_s/m_u$ have the same values 
when calculated from baryon masses and meson masses\cite{SakhZel,Gorky} 

The mass difference between $s$ and $u$ quarks calculated
in two ways from the linear term in meson and baryon masses showed
that it costs exactly the same energy to replace a nonstrange quark by a
strange quark in mesons and baryons, when the contribution from the
hyperfine interaction is removed.

\begin{equation}
\langle m_s-m_u \rangle_{Bar}= M_\Lambda-M_N=177\,{\rm MeV}
\end{equation}

\begin{equation}
\langle m_s-m_u \rangle_{mes} =
{{3(M_{K^{\scriptstyle *}}-M_\rho )
+M_K-M_\pi}\over 4} =180\,{\rm MeV}
\end{equation}

\begin{equation}
\left({{m_s}\over{m_u}}\right)_{Bar} =
{{M_\Delta - M_N}\over{M_{\Sigma^*} - M_\Sigma}} = 1.53
\eeq

\beq
\left({{m_s}\over{m_u}}\right)_{Mes} =
{{M_\rho - M_\pi}\over{M_{K^*}-M_K}}= 1.61
\end{equation}

Further extension of this approach led to two more relations for $m_s-m_u$ when
calculated from baryon masses and meson masses\cite{ICHJLmass,HJLMASS}. and to
three magnetic moment predictions with no free parameters\cite{DGG,Protvino}

\begin{equation}
\langle m_s-m_u \rangle_{mes} =
{{3 M_\rho + M_\pi}\over 8}
\cdot
\left({{M_\rho - M_\pi}\over{M_{K^*}-M_K}} - 1 \right)
= 178
\end{equation}
\begin{equation}
\langle m_s-m_u \rangle_{Bar}= 
{{M_N+M_\Delta}\over 6}\cdot
\left({{M_{\Delta}-M_N}\over
{M_{\Sigma^{\scriptstyle *}}-M_\Sigma}} - 1 \right)
=190.
\end{equation}
\begin{equation}
\mu_\Lambda=
-0.61
= \mu_\Lambda =
-{\mu_p\over 3}\cdot {{m_u}\over{m_s}} =
-{\mu_p\over 3} {{M_{\Sigma^*} - M_\Sigma} \over{M_\Delta - M_N}}
=-0.61 
\end{equation}

\begin{equation}
-1.46 =
{\mu_p \over \mu_n} =
-{3 \over 2}
\end{equation}

\begin{equation}
\mu_p+\mu_n= 0.88 
={M_{\scriptstyle p}\over 3m_u}
={2M_{\scriptstyle p}\over M_N+M_\Delta}=0.865 
\end{equation}
where masses are given in MeV and magnetic moments in nuclear magnetons.

\subsection{Problems in going beyond Sakharov-Zeldovich}

These successes and the success of the new relation (\ref{QQ1.7b}) make it 
difficult to improve on the results of the simple constituent quark model by
introducing new physics effects like higher order corrections. Any new effect
also introduces new parameters. In order to keep any analysis significant, it
is necessary to include large amounts of data in order to keep the total amount
of data much larger than the number of parameters. 

In contrast to the successes of the simple quark model in magnetic moments and
hyperon decay, there are also failures. Pinpointing these failures and comparing
them with the successes may offer clues to how to improve the simple picture. 

Combining the experimental data for hyperon magnetic moments and semileptonic
decays have provided some contradictions for models of hyperon structure.
The essential difficulty
is expressed in the experimental value of the quantity
\beq
{{(g_a)_{\Lambda \rightarrow p}}\over
{(g_a)_{\Sigma^-\rightarrow n}}}\cdot
{{\mu_{\Sigma^{+}} + 2\mu_{\Sigma^{-}}}\over{\mu_{\Lambda}}}=
0.12  \pm 0.04 
 \label{QQ1}  
\eeq
The theoretical prediction for this quantity from the standard SU(6)
quark model is unity, and it is very difficult to see how this
enormous discrepancy by a factor of $8 \pm 2$ can be fixed in any
simple way.
 
The expression (\ref{QQ1}) is chosen to compare two ways of
determining the ratio of the contributions of strange quarks
to the spins of the $\Sigma$ and $\Lambda$. In the commonly used
notation where $\Delta u (p) $, $\Delta d (p) $ and $\Delta s (p) $
denotes the contributions to the proton spin of the $u$, $d$ and
$s$ - flavored current quarks and antiquarks respectively to the spin of
the proton the SU(6) model gives

\beq 
\Delta s (\Lambda)_{SU(6)} = 1           \label{QQ2a} 
\eeq
\beq 
\Delta s (\Sigma)_{SU(6)} = -1/3         \label{QQ2b} 
\eeq
and
\beq 
{{\Delta s (\Sigma)_{SU(6)}}\over{\Delta s (\Lambda)_{SU(6)}}} =
{{(g_a)_{\Sigma^-\rightarrow n}}\over
{(g_a)_{\Lambda \rightarrow p}}} =
{{\mu_{\Sigma^{+}} + 2\mu_{\Sigma^{-}}}\over{3 \mu_{\Lambda}}}=
 -{{1}\over{3}} 
\label{QQ3a} 
\eeq
whereas experimentally
 \beq 
{{(g_a)_{\Sigma^-\rightarrow n}}\over
{(g_a)_{\Lambda \rightarrow p}}} =   -0.473 \pm 0.026
\label{QQ3b} 
\eeq
 \beq
{{\mu_{\Sigma^{+}} + 2\mu_{\Sigma^{-}}}\over
{3\mu_{\Lambda}}}= - 0.06 \pm 0.02 
\label{QQ3c} 
\eeq
The semileptonic decays give a value which
which is too large for the $\Sigma/\Lambda$ ratio; the magnetic moments
give a value which is too low. Thus the most obvious corrections to the
naive SU(6) quark model do not help. If they fix one ratio, they make
the other worse. Furthermore, the excellent agreement obtained by
De Rujula, Georgi and Glashow\cite{DGG}
 for $\mu_{\Lambda}$ assuming that the strange quark carries
the full spin of the $\Lambda$ suggests that eq. (\ref{QQ2a}) is valid, while
the excellent agreement of the experimental value $-0.340 \pm 0.017$
for $(g_a)_{\Sigma^-\rightarrow n}$ with the prediction -(1/3)
suggests that eq.(\ref{QQ2b}) is valid.
 
The disagreement sharpens the paradox of other disagreements previously
discussed because it involves only the properties of the $\Lambda$ and
$\Sigma$ and does not assume flavor SU(3) symmetry or any relation
between states containing different numbers of valence strange quarks.
There is also the paradox that the magnetic moment of the $\Lambda$ fits
the value predicted by the naive SU(6) quark model, while the magnetic
moments of the $\Sigma$ are in trouble. In the semileptonic decays it is
the opposite. It is the $\Sigma$ which fits naive SU(6) and both the
$\Lambda$ and the nucleon are in trouble. If one assumes the obvious fix
for the semileptonic decays by assuming a difference between constituent
quarks and current quarks, one can fit the nucleon and $\Lambda$ decays
but then the $\Sigma$ is in trouble.
 
The magnetic moments thus seem to indicate that the contribution of the
strange quark to the spin of the $\Sigma$ is smaller than any reasonable
model can explain, when the scale is determined by the $\Lambda$ moment.
This result is far more general than the simple naive SU(6) quark model.
But the new relation (\ref{QQ1.7b}) between the $\Sigma^+$ and $\Xi^o$ moments
seems to indicate that the strange quark contributions to these moments are
the same. 
  
\section{Semileptonic Decays} 

We now consider the semileptonic weak decays and begin by comparing the 
available data\cite{PDG}  for four semileptonic decays
with several theoretical predictions. 
The $\Xi^o \rightarrow \Sigma^+$ decay considered
above and equal to the neutron decay is omitted here.  

\pagebreak

{\centerline{\bf{TABLE 1. Theoretical Predictions and Experimental Values of
$G_A/G_V$}}}
$$ \vcenter{
\halign{${#}$\quad
        &${#}$\quad
        &${#}$\quad
        &${#}$\quad
        &${#}$\quad
        &${#}$\cr
DECAY & Simple &  Constituent & SU(3) & Experiment \cr
            &  SU(6)  & SU(6)& &    \cr
            &      &        &        &    \cr
n \rarrow p & 5/3 & input  & input
& 1.261 \pm 0.004\cr
            &      &        &        &    \cr
\Lambda \rarrow p & 1 & 0.756 \pm 0.003  & 0.727 \pm 0.007
& 0.718 \pm 0.015 \cr
            &      &        &        &    \cr
\Xi^- \rarrow \Lambda & 1/3 & 0.252 \pm 0.001  & 0.193 \pm 0.012
& 0.25 \pm 0.05 \cr
            &      &        &        &    \cr
\Sigma^- \rarrow n & -1/3 & 0.252 \pm 0.001  & input
& -0.340 \pm 0.017 \cr
{{\Sigma^- \rarrow n}\over
{\Lambda \rarrow p}}
& -1/3 & -1/3 & no\, prediction
& -0.473 \pm 0.026 \cr
}}   $$ \par
 
The nucleon and $\Lambda$ data are seen to be in strong
disagreement with simple SU(6) but are smaller by
about the same factor of about 5/4.
Thus they are both fit reasonably well by the
SU(6) constituent quark model which fixes $G_A/G_V$ for the constituent
quark to fit the nucleon decay data and reduces the other simple
SU(6) predictions by the same factor. But the $\Sigma$ data agree
with simple SU(6) and therefore disagree with constituent SU(6).
The SU(3) analysis fixes its two free parameters by using the nucleon
and $\Sigma$ decays as input; its predictions for the $\Lambda$ and
$\Xi$ fit the experimental data within two standard deviations.
However the error on the $\Xi$ data is considerably larger than
the other errors, and all three predictions fit the $\Xi$ data within two
standard deviations. Thus the significance of this fit can be questioned.
 
Our SU(3) fit deals directly with observable quantities rather than
introducing D and F parameters not directly related to physical
observables. This makes both the underlying physics and the
role of experimental errors much more transparent.
The neutron decay which has the smallest
experimental error fixes one of the two free parameters.
The $\Sigma^-$ decay provides the smallest error in fixing
the remaining parameter, the spacing between
successive entries in Table I, required to be equal by the SU(3) ``equal
spacing rule"  \cite{FRANKLIN}. The success of this procedure
is evident since the errors on the
predictions introduced by using these two decays as input are much
smaller than the experimental errors on the remaining decays.
 
The contrast between the good SU(6) fit of the $\Sigma$ and the bad SU(6)
fit of the others may give some clues to the structure of these baryons.
The $\Sigma$ data rule out the constituent SU(6) model which otherwise
seems attractive as it preserves all the good SU(6) results for strong and
electromagnetic properties at the price of simply renormalizing the axial
vector couplings to constituent quarks. Any success of SU(3) remains a puzzle
since no reasonable quark model has been proposed which breaks SU(6)
without breaking SU(3).

\section{The spin structure of baryons} 
\subsection{Results from DIS experiments}

Surprising conclusions about proton spin structure have arisen from an
analysis \cite{BEK}
combining data from polarized deep inelastic electron scattering and
weak baryon decays.
 
Polarized deep inelastic scattering (DIS) experiments provided 
high quality data for the spin structure functions of the proton, deuteron and
neutron~\cite{As88}-\cite{newpoldata}.
The first moments of the spin dependent
structure functions can be interpreted in terms of the contributions of
 the
quark spins ($\Delta \Sigma =\Delta u + \Delta d +\Delta s$) to the total spin
of the nucleon. The early EMC results 
\cite{As88} were very surprising, implying that 
$\Delta \Sigma$ is
 rather small
(about 10\%) and that the strange $sea$ is strongly polarized. 
More recent analyses \cite{Erice95,mk99}, 
incorporating higher-order QCD corrections,
together with most recent data,
suggest that
 $\Delta \Sigma$
is significantly larger, but still less than 1/3 of nucleon's helicity,
$\Delta \Sigma \approx 0.24 \pm 0.04$ and $\Delta s = -0.12 \pm 0.03$.

Conventional analyses to determine the quark contributions to
the proton spin, commonly denoted by $\Delta u $,
$\Delta d  $  and $\Delta s$, use use three
experimental quantities. The connection between two to proton spin 
structure is reasonably clear
and well established. The third is obtained from  
hyperon weak decay data  rather than nucleon data via 
SU(3) flavor symmetry relations and its use has been challenged. 

\subsection{How should SU(3) be used in analyzing hyperon decays
and relating data to baryon spin structure?}

We first note that the 
Bjorken sum rule together with isospin tell us that the neutron weak decay 
constant
\beq
g_A(n\rightarrow p) = \Delta u(p) - \Delta d(p) =  1.261 \pm 0.004
\eeq
and that
\beq
\Delta u(p) - \Delta d(p) =  \Delta d(n) - \Delta u(n) = 1.261 \pm 0.004
\label{QQ9a}
\eeq

Its SU(3) rotations give
\beq
g_A(\Sigma^-\rightarrow n) = \Delta u(n) - \Delta s(n) = -0.340 \pm 0.017 
\eeq
and 
\bea
\Delta u(n) - \Delta s(n) = \Delta d(p) - \Delta s(p) = 
 \nonumber \\
= \Delta s(\Sigma^-) - \Delta u(\Sigma^-) = -0.340 \pm 0.017 
\label{QQ9b}
\eea
as well as the prediction now  satisfied by experiment
\bea
g_A(\Xi^o \rightarrow \Sigma^+) = \Delta s(\Xi^o) - \Delta u(\Xi^o) =
 \nonumber \\
=  g_A(n\rightarrow p) =  1.261 \pm 0.004
\eea
The two independent linear combinations of $\Delta u(p)$, 
$\Delta d(p)$ and $\Delta s (p)$ obtained directly from the data without any 
assumptions about the $D$ and $F$ couplings commonly used 
can be combined to project out isoscalar component
of eq.(\ref{QQ9a}) and eq.(\ref{QQ9b}),
\bea
\Delta u +  \Delta d - 2 \Delta s = g_A(n\rightarrow p) + 
2g_A(\Sigma^-\rightarrow n) = 
 \nonumber \\ 
=0.58 \pm 0.03
\eea

The commonly used procedure to determine these two linear conbinations includes
the  data for the $\Lambda \rightarrow p$ and $\Xi^- \rightarrow \Lambda$
decays, which do not directly determine any linear combination of but require
an additional parameter, the $D/F$ ratio to give these quantities. Thus the
standard procedure uses includes two more pieces of data at the price of an 
additional free parameter. Since the $\Xi^- \rightarrow \Lambda$ decay has
a much larger error than all the other decays, there seems to be little point 
in introducing the D/F ratio.

\subsection{How does SU(3) symmetry relate the valence and sea quarks in the
octet baryons} 

We first note the following relations between the 
baryon spin structures following from SU(3) Symmetry 

 \bea 
\Delta u (p) = \Delta d (n) = \Delta u (\Sigma^+) = \Delta d (\Sigma^-) = 
 \nonumber \\
=\Delta s (\Xi^o) = \Delta s (\Xi^-) 
\label{QQ2.1a} 
   \eea
 \bea
\Delta d (p) = \Delta u (n) = \Delta s (\Sigma^+) = \Delta s (\Sigma^-) =
 \nonumber \\
= \Delta s (\Sigma^o) = \Delta u (\Xi^o) = \Delta d (\Xi^-) 
\label{QQ2.1b}    
\eea
\bea
\Delta s (p) = \Delta s (n) = \Delta d (\Sigma^+) = \Delta u (\Sigma^-) = 
 \nonumber \\
=\Delta d (\Xi^o) = \Delta u (\Xi^-) \label{QQ2.1c}    \eea
 \beq 
\Delta u (\Sigma^o) = \Delta d (\Sigma^o) = (1/2)\cdot 
[\Delta u (\Sigma^+) + \Delta d (\Sigma^+) ]  
\label{QQ2.2a}    
\eeq
 \beq 
\Delta q (\Sigma^o) + \Delta q (\Lambda) = (2/3)\cdot 
[\Delta u (n) + \Delta d (n) + \Delta s (n) ]  
\label{QQ2.2b}    
\eeq

These relations allow all the baryon spin structures to be obtained from the
values of $\Delta u (n)$, $\Delta d (n) $  and $\Delta s (n) $

However, we know that SU(3) symmetry is badly broken. This can be seen easily 
by noting that all these SU(3) relations apply separately to the valence 
quark and sea quark spin contributions. Thus SU(3) requries that the sea 
contributions satisfy eq.(\ref{QQ9b}).

Since the strange contribution of the sea in the proton is known experimentally
to be suppressed\cite{CCFR}, this suggests that the strange sea in the 
$\Sigma$ must be enhanced. This simply does not make sense in any picture 
where SU(3) is broken by the large mass of the strange quark. We are therefore 
led naturally to a model in which SU(3) symmetry holds for the valence quarks
and is badly broken in the sea while the sea is the same for all octet baryons, 
is a spectator in weak decays and does not contribute to the magnetic 
moments. The sea thus does not contribute to the coupling of the photon or the
charged weak currents to the nucleon. The one place where the sea contribution 
is crucial is in the DIS experiments, which measure the coupling of the neutral 
axial current to the nucleon. 

The Bjorken sum rule and its SU(3) rotations relate the weak decays to the 
spin contributions of the active quarks to the baryon, without separating them
into valence and sea contributions. The effects of the flavor symmetry breaking
in the sea can be avoided by assuming that the flavor symmetry is exact for
the algebra of currents, but the the hadron wave functions are not good SU(3) 
states but are broken in the sea. In this way one can obtain relations for the
differences between spin contributions in which the sea contribution cancels
out if the sea is the same for all octet baryons, even if SU(3) is broken
in the sea. 

\section{Where is the physics? What can we learn?}
  \subsection{How is SU(3) broken?}

We now examine the underlying physics of some of these decays in more
detail. The weak
decays measure charged current matrix elements, in contrast to the
EMC experiment which measures neutral current matrix elements
related directly via the Bjorken sum rule to
$\Delta u (p) $, $\Delta d (p) $ and $\Delta s (p) $.
The charged and neutral current matrix elements have been related
by the use of symmetry assumptions whose validity has been questioned
\cite{LIP,HJL}.

We now examine the $\Sigma^- \rightarrow n$ decay and see how SU(3) breaking
affects the relations 
\beq
 g_A(\Sigma^- \rightarrow n)  =  \Delta u(n) -\Delta s(n)=
 \Delta d(p) -\Delta s(p)
\label{eIIIa} 
 \eeq 
\beq
 \Delta s(\Sigma^-) -\Delta u(\Sigma^-) = \Delta d(p) -\Delta s(p)
 \label{eIIIb} 
  \eeq 
\beq
 |G_V(\Sigma^- \rightarrow n)|  = |G_V(n \rightarrow p)|
 \label{eIIIc}
 \eeq

The quantity denoted by $g_A$ is a
ratio of axial-vector and vector matrix elements. Although only the axial
matrix element is relevant to the spin structure,
breaking SU(3) in the baryon wave functions breaks both the 
relations between axial and vector couplings,
as well as those from CVC for strangeness changing currents. 
Serious constraints on possible SU(3) breaking
in the baryon wave functions are placed by the known agreement 
with Cabibbo theory of experimental 
vector matrix elements,
uniquely determined in the SU(3) symmetry
limit.  On the other hand, the strange quark
contribution to the proton sea is already known from experiment
to be reduced roughly by a factor of two from that of a flavor-symmetric
sea \cite{CCFR}, due to the effect the strange quark mass. 
This suppression is expected to violate the $\Sigma^- \leftrightarrow n$ 
mirror symmetry, since it is hardly likely that the strange sea should
be enhanced by a factor of two in the $\Sigma^-$. Yet Cabibbo theory
requires retaining the relation between the vector matrix elements
eq.(\ref{eIIIc}).

 \subsection{A model which breaks SU(3) only in the sea}

We now move to the discussion of the model described above mechanism for 
breaking SU(3) \cite{PJECH} which
keeps all the good results of 
Cabibbo theory like 
eq.(\ref{eIIIc}) by introducing
a baryon wave function 
\beq 
\ket{B_{phys}}= \ket{B_{bare} \,\cdot \phi_{sea}(Q=0)} 
\label{WW1.2}
\eeq
where $\ket{B_{bare}}$ denotes a valence quark wave function which is an 
SU(3) octet satisfying the condition eq.(\ref{eIIIc})
and $\phi_{sea}(Q=0)$ denotes a sea with zero electric 
charge which may be flavor asymmetric but is {\em the same} for all 
baryons. 
The wave function eq.(\ref{WW1.2}) is shown\cite{numagmom}  
to satisfy eq.(\ref{eIIIc}) and to 
{\em to give all
charged current matrix elements by the valence quark component.} This provides
an explicit justification for the hand-waving argument \cite{PJECH} that the 
sea behaves as a spectator in hyperon decays.

Unlike the charged current,
the matrix elements of the neutral components of the weak currents 
{\em do} have
sea contributions, and these contributions are observed in the DIS experiments.
The SU(3) symmetry relations eqs.(\ref{QQ2.1a}-\ref{QQ2.2b}) are no longer 
valid. 
However, the weaker relation obtained from current algebra \cite{HJL94}
still holds.
\beq 
g_A(\Sigma^-\rightarrow n) =
{{\bra {n}  \Delta u - \Delta s \ket {n}  -
\bra {\Sigma^-}  \Delta u - \Delta s \ket {\Sigma^-}}
\over{2}}
\label{WW1.6}
\eeq
SU(3) says 
\beq 
{\bra {\Xi^o}  \Delta s - \Delta u \ket {\Xi^o}  =
\bra {p} \Delta u - \Delta d \ket {p}}
\label{WW9.6}
\eeq
If the strange sea is suppressed, this is clearly wrong.
However, Current Algebra relations require only that 
\bea 
\bra {\Xi^o}  \Delta s - \Delta u \ket {\Xi^o}  +
\bra {\Sigma^+}  \Delta u - \Delta s \ket {\Sigma^+} =
\nonumber \\  
= \bra {p}  \Delta u - \Delta d \ket {p}  +
\bra {n}  \Delta d - \Delta u \ket {n}  
\label{WW9.7}
\eea

This is immune to strange sea suppression in all baryons. 

 \subsection{Getting $\Delta u $, $\Delta d $  and $\Delta s $
 From Data }

Breaking up the quark contributions into valence and sea contributions
becomes necessary to treat  SU(3) breaking and the suppression of the strange 
sea.
Two ways of doing this have been considered\cite{numagmom}, one using hyperon
decay data and the other using the ratio of the proton and neutron magnetic
moments. 
 
What is particularly interesting is that each of the two approaches makes
assumptions that can be questioned, but that although these assumptions are
qualitatively very different, both give very similar results. 
The use of hyperon data
requires a symmetry assumption between nucleon and hyperon wave functions, which
is not needed for the magnetic moment method. But the use of magnetic moments
requires that the sea contribution to the magnetic moments be negligible, which
is not needed for the hyperon decay method.

\section{Conclusions}

The question how flavor symmetry is broken remains open. Model builders must 
keep track of how proposed SU(3) symmetry
breaking effects affect the good Cabibbo results for hyperon decays
confirmed by experiment.  The observed
violation of the Gottfried sum rule remains to be clarified, along with
the experimental question of whether this violation of $\bar u - \bar
d$ flavor symmetry in the nucleon exists for polarized as well as for
unpolarized structure functions. The question of how SU(3) symmetry is
broken in the baryon octet can be clarified by experimental measurements
of $\Lambda$ polarization in various ongoing experiments.

\acknowledgments
It is a pleasure to thank Danny Ashery, Marek Karliner and Joe Lach  
for helpful discussions and comments. This work
was partially supported by a grant from US-Israel Bi-National Science 
Foundation.
 
{
\tighten

}

\end{document}